\documentclass[prl,twocolumn,showpacs,superscriptaddress]{revtex4}% Physical Review B
\usepackage{graphicx}% Include figure files
\usepackage{dcolumn}% Align table columns on decimal point
\usepackage{bm}% bold math

\begin{document}
\title{Magnetic fluctuations from stripes in cuprates}%Force line breaks with \\
\author{G. Seibold}
\affiliation{Institut f\"ur Physik, BTU Cottbus, PBox 101344, 
         03013 Cottbus, Germany}
\author{J. Lorenzana}
\affiliation{Center for Statistical
Mechanics and Complexity, INFM, Dipartimento di Fisica,
Universit\`a di Roma La Sapienza, P. Aldo Moro 2, 00185 Roma, Italy}
\date{\today}% It is always \today,
\begin{abstract}
Within the time-dependent 
Gutzwiller approximation for the Hubbard model we compute the 
magnetic fluctuations of vertical metallic stripes 
with parameters appropriate for  La$_{1.875}$Ba$_{0.125}$CuO$_4$ (LBCO).
For bond- and site-centered 
stripes the excitation spectra are similar, consisting of a low-energy 
incommensurate acoustic branch which merges into
a  ``resonance peak'' at the antiferromagnetic wave vector
%$(\pi/a,\pi/a)$
 and several high-energy 
optical branches. The acoustic  branch is similar to the result of 
theories assuming localized spins whereas the
optical branches are significantly different.
Results are in good agreement
with a recent inelastic  neutron study of LBCO. 
%in a wide energy and momentum range.
%establishing stripes as a theoretically
%well founded phenomenon.
\end{abstract}
\maketitle
Since the discovery of cuprate superconductors the elucidation of their
magnetic properties has been a subject of intense research in the
high-$T_c$ community due to their possible relevance for the
superconducting mechanism\cite{kas98}.  
However, no consensus has been reached yet wether the magnetic 
excitation spectra in the various cuprate materials can be traced back
to some universal phenomenology which could be expected in the face of the
robust nature of superconductivity.
The insulating parent compounds
show long-range antiferromagnetic (AF) order in the CuO$_2$ planes
below the N{\'e}el temperature\cite{man91}. This static AF order is lost above
a concentration of added holes per planar copper $x\approx 0.02$ but complex dynamical 
spin correlations persist up to the overdoped regime\cite{kas98,yam98}.

Regarding the lanthanum cuprates (LCO), neutron scattering (NS) experiments
have revealed low-energy incommensurate magnetic excitations
at wave-vectors (hereafter in units of $2\pi/a$\cite{units})  $(1/2\pm\epsilon,1/2)$ and
$(1/2,1/2\pm\epsilon)$\cite{swche91}  in the doping regime where the
material is superconducting ($x>0.05$).
The incommensurability $\epsilon$ depends linearly on 
doping $\epsilon=x$ up to $x \approx 1/8$ and stays constant
beyond\cite{yam98}.
This behavior can be successfully explained\cite{fle01b,lor02b,sei04a}
by metallic, self-organized quasi-one-dimensional structures called 
stripes\cite{zaa89mac89hsch90poi89}
which are oriented parallel to the Cu-O bond (hereafter vertical-stripes)
and act as antiphase domain walls for the antiferromagnetic order.
The strongest experimental support for the stripe scenario  
stems from neutron\cite{tra95,fuj04} and $x$-ray scattering\cite{nie99} 
experiments on LCO where co-dopants induce a pinning
of the stripes so that the associated charge order can be detected.

In YBa$_2$Cu$_3$O$_{6+y}$  (YBCO) incommensurate magnetic fluctuations have 
 been detected by inelastic NS
(INS)\cite{moo98,ara99,dai01,hay04} with a similar doping dependence 
of the low-energy
incommensurability to that of LCO\cite{dai01}.
 Upon increasing energy  the incommensurate branches
continuously disperse towards the so-called resonance mode at
wave-vector $Q_{AF}=(1/2,1/2)$, a  collective magnetic mode that grows
up below $T_c$\cite{ros91,fon95,pbou96}.  The energy $E_r$ of the spin
resonance seems to scale linearly with $T_c$ which has led to speculations
that it could be related to the superconducting pairing and phase coherence 
(cf. Ref.~\cite{dai99,dai00} and references therein), thus 
suggesting a magnetic origin of the high transition temperatures of cuprates.
Finally above $E_r$ the magnetic fluctuations in YBCO acquire again an 
incommensurate structure\cite{ara99,ara00,rez03,moo02,hay04}
and are already observed above $T_c$\cite{hay04}.

Very recent experiments were dedicated to explore 
the problem of universality in the magnetic excitations between
different cuprate families\cite{tra04,rez03,pai04,chr04}.
In particular Tranquada and collaborators\cite{tra04} reported INS measurements
of the magnetic excitations in La$_{1.875}$Ba$_{0.125}$CuO$_4$ 
which shows static charge {\it and } spin order.
 Most interestingly the dispersion
of spin excitations shows features which resemble closely those of
YBCO although the measurement has been performed above $T_c$. 
The similarity between both compounds has been further demonstrated
 by a high energy study of YBCO\cite{hay04} and a high resolution experiment 
on optimally doped LCO\cite{chr04}.
These works\cite{tra04,hay04,rez03,pai04,chr04} are
certainly an important step towards a unified understanding of
magnetic fluctuation in cuprates. On the other hand they open new
questions. 
The spin excitations in YBCO are commonly explained 
in terms of an itinerant picture as arising from
a dispersing two-particle bound state induced by AF correlations
in a $d$-wave superconducting system\cite{bri99yin00nor00onu02,pai04}.
On the other hand magnetic excitations on top of stripes are usually 
described in a localized moment picture within spin-wave 
theory (LSWT)\cite{bat01kru03car04,boo03}. 
If the origin of magnetic excitations in cuprates is universal, 
what picture is more appropriate? 
The spectra of LSWT 
are difficult to reconcile with several features of 
 YBCO\cite{hay04,pai04,rez03} and LCO\cite{tra04,chr04} challenging 
the stripe interpretation itself\cite{rez03,chr04}.  
Here we show that the magnetic excitations in
La$_{1.875}$Ba$_{0.125}$CuO$_4$ (and possibly in the other cuprates) 
can be understood in terms of the spin fluctuations of metallic
stripes which are in an intermediate regime,
i.e. neither the localized nor the itinerant picture
applies. Contrary to the LSWT computations\cite{bat01kru03car04}
our results are in agreement with experiment\cite{tra04}
 over the whole range of energies  and momenta.

Our investigations are based on the one-band Hubbard model with
hopping restricted to nearest ($\sim t$) and next nearest ($\sim t'$)
neighbors
\begin{displaymath}
H=-t\sum_{\langle ij \rangle,\sigma}c_{i,\sigma}^{\dagger}c_{j,\sigma}
- t'\sum_{\langle\langle ij\rangle\rangle,\sigma}c_{i,\sigma}^{\dagger}
c_{j,\sigma}
+ U\sum_{i}
n_{i,\uparrow}n_{i,\downarrow}.\nonumber
\end{displaymath}
Here $c_{i,\sigma}^{(\dagger)}$ destroys (creates) an electron
with spin $\sigma$ at site
$i$, and $n_{i,\sigma}=c_{i,\sigma}^{\dagger}c_{i,\sigma}$. $U$ is the
on-site Hubbard repulsion.

Static charge and spin textures are obtained within an unrestricted Gutzwiller
approximation (GA). Dynamical properties are computed on top of the
inhomogeneous solutions within
the time-dependent GA\cite{sei01} (TDGA) which previously has 
been shown to provide an accurate description of the optical conductivity 
within the more realistic three-band model\cite{lor03}. Here, because we restrict to 
low-energy magnetic excitations, we do not expect that interband 
effects will play an important role and therefore a one-band
description should be sufficient. 

Parameters were fixed by requiring that a) the linear concentration
of added holes is $1/(2a)$ according to
experiment\cite{yam98,tra95,tra04} and b) a TDGA computation of the
undoped AF insulator reproduces the experimental dispersion
relation\cite{col01}. Condition a) was shown to be very sensitive to
$t'/t$ \cite{sei04a} whereas condition b) is sensitive to $U/t$ and $t$, the former
parameter determining the observed energy splitting between magnons  
at wave-vectors $(1/2,0)$ and $(1/4,1/4)$\cite{col01}. Indeed, the
splitting vanishes 
within spin-wave theory applied to the Heisenberg model
which corresponds to $U/t \rightarrow \infty$. We find that both
conditions are met by  $t'/t=-0.2$, $U/t=8$ and
$t=354$meV. In this way all parameters are fixed and
the subsequent computation of the magnetic fluctuation of the stripes can
be considered without free parameters.
Results shown below are obtained 
in a $40\times 40$ sites lattice and for doping $x=1/8$
corresponding to a period of charge modulation  $d=4a$.
\begin{figure}[tbp]
\includegraphics[width=8cm,clip=true]{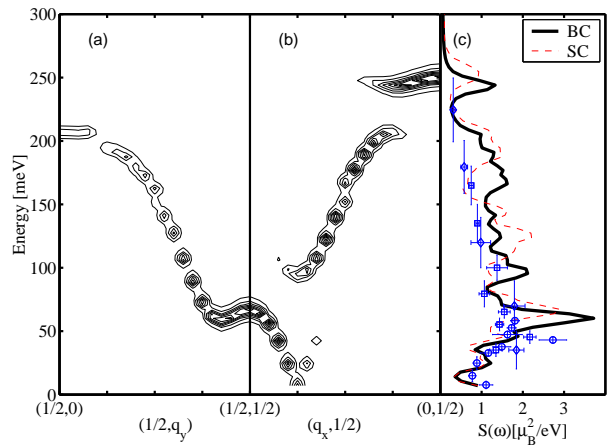}
\caption{10 contour levels of $\omega S^{\perp}(\omega,{\bf q})$ at integer values showing the
  dispersion of magnetic excitations along  (a) and perpendicular
  (b) to an array of BC stripes. The factor of  $\omega$ serves to cancel an
  infrared intensity divergence. Panel (c) shows 
  $S(\omega,{\bf q})$ integrated on the magnetic Brillouin
  zone around (1/2,1/2)\cite{personal} for BC and SC stripes together with the
  experimental data\cite{tra04}. 
For the meaning of ``error'' bars see Ref.~\cite{tra04}. 
}
\label{front} 
\vspace*{-0.5cm}
\end{figure}
 The charge and spin structure of the static GA solution 
is similar to the one reported in Ref.~\cite{sei04a} where a 
slightly different
parameter set has been used. The local magnetization of BC stripes
is reduced to 0.20 (0.38) with respect to 1/2 inside (outside) 
the core of the charge domains. The band structure is quasi-one dimensional 
with a Fermi momentum  $k_F\sim 1/8$.  Bond-centered (BC) and
site-centered (SC) stripes are almost degenerate in energy which is also 
found in other one-band calculations \cite{whi98prl80,fle01b} and within
 the three-band model at higher doping\cite{lor02b,lor03}. 
We anticipate that for both structures  the magnetic excitation
spectra are rather similar 
[see Fig.~\ref{front}(c)]
with differences regarding the intensity distribution 
and gaps between magnetic bands.
For definiteness we mainly restrict ourself to the BC case because 
of marginally better agreement with experiment (see below). 
Additionally these textures constitute 
the more stable configuration at $x=1/8$  in the more accurate 
three-band model\cite{lor02b} and in first principle computations\cite{ani04}.

We compute the transverse dynamical structure factor, 
$S^{\perp}(\omega,{\bf q})=\frac{(g\mu_B)^2\eta Z_{\chi}}{N\hbar}\sum_{\nu} 
|\langle0|S^+_{\bf q}|\nu\rangle|^2 \delta(\omega-\omega_{\nu})$
which is probed by INS. Here $g=2$ and as in LSWT we included the renormalization 
factor $Z_{\chi}$\cite{man91}. In the insulating phase our spectral weights are close 
to LSWT and therefore we adopt $Z_{\chi} = 0.51$\cite{col01}.
 $\eta$ takes into account polarization factors in the NS
  cross section and will be discussed below.
 Energy is sampled at intervals of 3.5meV.
 
A contour level plot of $\omega S^{\perp}(\omega,{\bf q})$ [Fig.~\ref{front}(a),(b)]
shows the dispersion of magnetic excitations for stripes
oriented along the $y$-axis.  
The lower (acoustic) branch perpendicular to the stripes (b) 
is similar to the lowest branch found within LSWT\cite{bat01kru03car04}.
Indeed, since the acoustic branch involves long-wavelength excitations
it should not depend on the short-range details of the model.
It shows the correct Goldstone-like behavior going to zero frequency at the
ordering wave-vector $Q_s=(1/2\pm\epsilon,1/2)$ with $\epsilon=1/8$.
Starting from $Q_s$ one observes two branches of spin-waves where the
one dispersing towards smaller $q_x$  rapidly loses intensity. 
The other one remains very intense up to 
 $Q_{AF}$  where it can be associated with the resonance peak. 
Moreover, as shown in  Fig.~\ref{front}(a,b) the dispersion 
develops a local maximum at $Q_{AF}$ explaining the strong intensity
in the momentum integrated structure factor [Fig.~\ref{front}(c)] at
the resonance frequency.  In the direction
 of the stripe the excitations display a ``roton-like'' minimum. 
The energy of the resonance at $Q_{AF}$ is $E_{r}=65$meV which is 
reasonably close to the experimental one for this system
$E_{r}=55$meV\cite{tra04}  
taking into account the simplicity of the
model and the fact  that we have no free parameters.  
% Also the ``roton-like'' minimum contributes significantly to the
%main peak (on the low energy side) in the momentum integrated
%structure factor [Fig.~\ref{front} (c)]. 

In order to compare with experiments one should average 
over the two possible orientations of the stripes. In  Fig.~\ref{front} 
this amounts to adding to each of the panels (a) and (b) the data of 
the other panel reflected with respect to the central axes. One obtains 
 an X-shaped dispersion providing a natural explanation 
for the X-like feature seen in both YBCO\cite{ara99,moo02,pai04} 
and LCO\cite{tra04}. 

The acoustic branch and its continuation in Fig.~\ref{front}(a) are
quite similar to the dispersion obtained in a 
weakly coupled two-leg ladder system\cite{voj04uhr04}. 
In these approaches parameters are adjusted to drive the 
system into the quantum critical point separating the 
quantum paramagnet from the magnetically ordered state\cite{sac03}. 
Hence the similarity with our spectra is reasonable 
since regarding the magnetism 
the ordered state corresponds to the present ground state 
and one can expect continuity of the excitations at the 
transition. The spin-leg ladder theories\cite{voj04uhr04}, however,
rely heavily on the fine tuning of  the coupling parameter between the legs 
and an even charge periodicity 
of the stripes (like for $\epsilon=1/8$). In contrast we obtain qualitatively 
similar spectra for  $\epsilon=1/8$ and $\epsilon=1/10$ 
($x=0.1$, $d=5a$)\cite{lor02b,sei04a}, in accord
with experiment\cite{chr04}. Note that this does not exclude the
interesting possibility that a small spin-gap 
opens by the ladder mechanism for $x\ge
1/8$ and $\epsilon=1/8$\cite{sac03,ani04,voj04uhr04}.

Above the acoustic branch in Fig.~\ref{front}(b) there are three optical
branches. The lowest two almost touch at $q_x=1/4$. Gap positions 
are in agreement with LSWT but the dispersion is not. Indeed the
computations in Ref.~\cite{bat01kru03car04}
yield a shift of the two lower optical branches by 1/8 along $q_x$ with
respect to our computation. 
Formally this difference can be traced back to the non local
nature of magnetic excitations in our system which shows
magnetic moments considerably smaller than in the insulator, and 
more importantly is metallic. In LSWT
only processes of the kind $S^+_iS^-_j$ are allowed in the
effective interaction kernel with $i$ and $j$ being close neighbors\cite{bat01kru03car04}). 
We find that not only processes with $i$ far from $j$ are important but
also processes of the kind 
$c_{i,\uparrow}^{\dagger}c_{i',\downarrow}c_{j,\downarrow}^{\dagger}c_{j',\uparrow}$
with all four sites different.

The absolute spectral weights are determined by the parameter 
$\eta$. If one assumes a three-dimensional isotropic distribution of 
magnetic domains $\eta=2/3$. If instead the magnetization is parallel
to the CuO plane $\eta$ will decrease with energy and 
take values closer to 1/2. We adopt for $\eta$  80\% of the isotropic 
value which gives good account of the lower part of the spectra. (At
higher energies a further decrease is expected \cite{personal}). 
For a slightly more doped sample\cite{hay96} we find $\eta\sim 2/3$.

Some details of the relative spectral weight are better reproduced
 by BC stripes. {\it i}) 
The resonance energy $E_r$ is closer to the experimental value in
the BC case. {\it ii}) 
For BC stripes the gap between the acoustic branch and the 
first optical branch
produces a dip in the integrated intensity at 80meV followed by a peak
at 100meV which is in agreement with
experiment contrary to the SC case. Although these features favor BC 
stripes more theoretical and experimental
work is needed to confirm this finding. 
\begin{figure*}[tbp]
\hspace*{-1cm}\includegraphics[width=18cm]{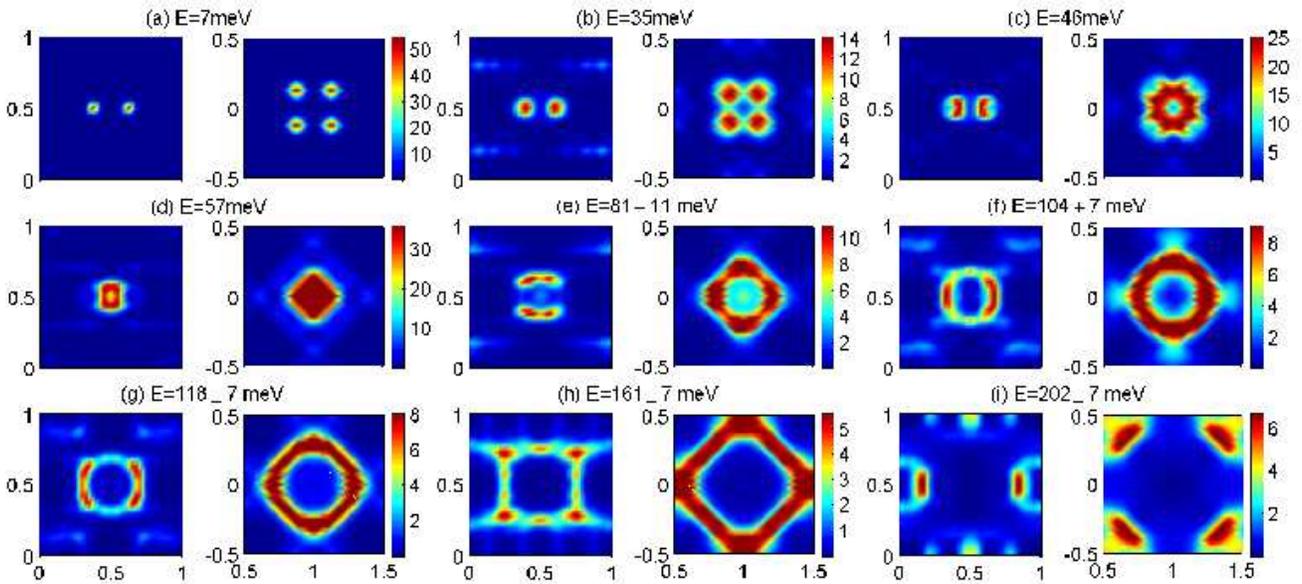}
\caption{Constant frequency scans of $S^{\perp}(\omega,{\bf q})$ for BC
  stripes convolved  with a Gaussian (FWHM=0.06 r.l.u.).
The title indicates the energy or the energy window (e-i)
over which intensities are averaged.  
For each figure (a)-(i) the left panel shows the
intensity for stripes oriented along the y-direction (divided by 2).
 In order to
compare with the experimental data of Ref.~\protect\onlinecite{tra04}
the right panel shows the
average of both horizontal and vertical stripes with  the coordinate
system rotated by $45^0$ with wave-vector units of
$2\pi/(\sqrt{2}a)$. Parameters are as in Fig.~\ref{front} and as
explained in the text.}
\label{cont}
\vspace*{-0.5cm}
\end{figure*}

Fig.~\ref{cont} reports constant frequency scans of 
$S^{\perp}(\omega,{\bf q})$.
At the lowest energy for y-axis oriented stripes [panel (a) left] one
intersects the Goldstone mode close to the 
incommensurate wave-vector  $Q_s$. At higher energy [(b),(c)]  one
intersects the spin-wave cone. The intensity is very anisotropic along
the intersection and only the region closest to $Q_{AF}$ is visible.
This effect is also apparent from Fig.~\ref{front}. 
Although our study focusses on an underdoped system 
the large intensity difference fits nicely with the fact that 
 only the branch closest to $Q_{AF}$ is seen in a recent high resolution 
study  on optimally doped LCO\cite{chr04}. Immediately below the resonance
[Fig.~\ref{cont} (d)] the spin excitations have shifted close to
$Q_{AF}$ and one observes the appearance of intensity
in the stripe direction due to the ``roton-like'' behavior. 
At 81meV one intersects the first gap of Fig.~\ref{front} (b) so that the
main features are due to excitations along the stripes. At higher energies the
contributions from the modes perpendicular to the stripe become
gradually more important. 
Moreover, the orientation averaged scans are in remarkably good
agreement with the corresponding panels of Ref.~\cite{tra04}. 

Another very interesting finding are the horizontal cigar like
features seen at 35meV and higher energies. Similar features have been
obtained in weak coupling in the longitudinal channel\cite{kan01}. 
These structures  
appear close to $q_y\sim 1/4 \sim 2k_F$ and correspond to  
spin-flip  backscattering processes
of the quasi-one dimensional metallic subsystem. 
  At this $q_y$ the
continuum extends to about 100meV.  As expected
there is also a sharp collective mode corresponding to forward
scattering processes but
with too weak intensity to be observable in Fig.~\ref{cont}. 
Diffusive scattering revealed a similar phenomenon
in La$_{5/3}$Ni$_{1/3}$O$_4$ where stripes are
insulating but have a spin-1/2 degree of freedom at the core\cite{boo03prl}.

Concluding, we have calculated the magnetic excitations of cuprates in the 
stripe phase within the TDGA applied to an extended Hubbard model.
From a methodological point of view the present work together with
our previous study Ref.~\cite{lor03} shows that 
TDGA is accurate and simple enough in order to obtain realistic dynamical 
properties of textured strongly correlated system. 
Stripes in our computation are metallic charge and spin density-waves in a
regime intermediate between localized moment and itinerant pictures. 
Our results are in good agreement with experiments in  
La$_{1.875}$Ba$_{0.125}$CuO$_4$ providing a straightforward
explanation of the energy and momentum dependent evolution of 
$S^{\perp}(q,\omega)$ in terms of stripes
 and hence confirming stripes
as a robust feature of this system with a firm theoretical basis.
The fact that 
the magnetic excitation spectra are similar in different cuprate
families and for different doping levels suggests that stripes 
or the proximity to stripe instabilities are a universal property of
cuprates and hence that they may be relevant for the superconducting
mechanism. 

We acknowledge invaluable insight from J. M. Tranquada, A. T. Boothroyd
and R. Coldea.
G. S. acknowledges financial support from the Deutsche Forschungsgemeinschaft.
 

\begin{thebibliography}{10}
\bibitem{units} $a\sim3.8$\AA~ is the 
lattice spacing in tetragonal coordinates.
\end{thebibliography}
 \end{document}